\documentclass[useAMS,usenatbib,fleqn]{mn2e}
\usepackage{listings}
\usepackage{amsmath}
\usepackage{natbib}
\usepackage{graphicx}
\usepackage{color}
\usepackage{rotating}
\usepackage{nicefrac}
\usepackage{txfonts}
\usepackage{subfigure}
\usepackage{trivfloat}
\trivfloat{listfloat}
\usepackage{longtable}
\usepackage[percent]{overpic}
\usepackage{xspace}
\usepackage{url}
\usepackage{mathrsfs}
\usepackage{amssymb}
\usepackage{mathtools} 
\usepackage{enumitem}
\usepackage[percent]{overpic}
\usepackage{soul}

\usepackage{epsfig}
\usepackage{amsmath}
\usepackage{rotating}
\usepackage{natbib}
\usepackage{multirow}
\usepackage[noend]{algpseudocode}

\newcommand{\sima}{\textit{Simulation Archive}\xspace}
\newcommand{\simas}{\textit{Simulation Archives}\xspace}
\newcommand{\reb}{{\sc \tt REBOUND}\xspace}
\newcommand{\whfast}{{\sc \tt WHFast}\xspace}
\newcommand{\ias}{{\sc \tt IAS15}\xspace}

\lstdefinestyle{customc}{
  belowcaptionskip=1\baselineskip,
  breaklines=true,
  language=C,
  showstringspaces=false,
  basicstyle=\footnotesize\ttfamily,
}

\usepackage{array}
\usepackage{appendix}
\usepackage{comment}

\def \aap{A\&A}

\def \apjl{ApJ}
\def \apjs{ApJS}

\def\gsim{\;\rlap{\lower 2.5pt
 \hbox{$\sim$}}\raise 1.5pt\hbox{$>$}\;}
\def\lsim{\;\rlap{\lower 2.5pt
   \hbox{$\sim$}}\raise 1.5pt\hbox{$<$}\;}

\title{A new paradigm for reproducing and analyzing N-body simulations of planetary systems}
\date{Accepted 2017 January 24. Received 2017 January 20; in original form 2016 November 18}

\author[Hanno Rein, Daniel Tamayo]{ 
Hanno Rein$^{1,2}$ and Daniel Tamayo$^{1,3,4}$  
\\
$^1$ Department of Physical and Environmental Sciences, University of Toronto at Scarborough, Toronto, Ontario M1C 1A4, Canada\\
$^2$ Department of Astronomy and Astrophysics, University of Toronto, Toronto, Ontario, M5S 3H4, Canada\\
$^3$ Canadian Institute for Theoretical Astrophysics, 60 St. George St, University of Toronto, Toronto, Ontario M5S 3H8, Canada\\
$^4$ Centre for Planetary Sciences Fellow
}

\vspace{0.5\baselineskip}

\begin{document}
\maketitle

\begin{abstract}
The reproducibility of experiments is one of the main principles of the scientific method.
However, numerical N-body experiments, especially those of planetary systems, are currently not reproducible.
In the most optimistic scenario, they can only be replicated in an approximate or statistical sense.
Even if authors share their full source code and initial conditions, differences in compilers, libraries, operating systems or hardware often lead to qualitatively different results.

We provide a new set of easy-to-use, open-source tools that address the above issues, allowing for exact (bit-by-bit) reproducibility of $N$-body experiments.
In addition to generating completely reproducible integrations, we show that our framework also offers novel and innovative ways to analyze these simulations.
As an example, we present a high-accuracy integration of the Solar System spanning 10~Gyrs, requiring several weeks to run on a modern CPU.
In our framework we can not only easily access simulation data at predefined intervals for which we save snapshots, but at any time during the integration.
We achieve this by integrating an on-demand reconstructed simulation forward in time from the nearest snapshot.
This allows us to extract arbitrary quantities at any point in the saved simulation exactly (bit-by-bit), and within seconds rather than weeks.

We believe that the tools we present in this paper offer a new paradigm for how $N$-body simulations are run, analyzed, and shared across the community.

\end{abstract}


\section{Introduction}
\label{sec:intro}
Scientists have studied the motion of gravitationally interacting bodies such as the planets in the Solar System for centuries \citep[for a comprehensive historical review see][]{Laskar2012}.
Today, such studies are mostly done on computers through $N$-body experiments.
Unsurprisingly, answering new questions often requires pushing the limits of current technology. 
When studying the dynamical evolution of astrophysical systems, this often translates to several months of computing time.

Working with such simulations can be cumbersome.
Traditionally, one has to specify both the quantities to record and the output cadence ahead of time.  
But this is not easy to do, as the appropriate choices are often only clear {\it after} an initial analysis of the data.
The problem is compounded if one considers that other groups (perhaps with different goals) might also want to analyze the output data at a later time.

Even worse, current N-body integrators yield trajectories that are not individually reproducible.   
The central problem is that most dynamical systems of interest (e.g., the Solar System) are chaotic.
Thus, any discrepancy in the representation of floating point numbers amplifies exponentially fast.  
This means that seemingly insignificant differences in hardware, software, initial conditions, or even the choice of which times to output data can change, e.g., whether two planets collide or not in a given simulation.

For example, \cite{LaskarGastineau2009} showed by direct integration that in about~$1\%$ of their simulations, Mercury collides with another terrestrial planet or plunges into the Sun in the next~5~Gyr.  
Since then, several other groups have studied the physical origin of this instability \citep{LithwickWu2011,Batygin2015}, but could not directly compare their theory to the $6.2$~million CPU-hour dataset produced by \cite{LaskarGastineau2009}.
As we explain further below, if one is interested in analyzing the simulations resulting in collision, it is not enough to have the same code and exact initial conditions.
One would have to expend comparable computational time to generate many simulations, and select the (different!) ones that lead to collisions.

Chaos in N-body integrations therefore not only leads to inefficient use of computational resources, it is also a fundamental challenge to reproducibility and a barrier to the scientific process.
In this paper we present solutions to the technical obstacles hampering reproducibility, and propose a \textit{new paradigm} for performing N-body integrations.
In our framework, machine-independent integration algorithms generate reloadable binary snapshots of the simulation as it progresses.
The resulting \sima can then be shared across the community, and different members can robustly choose {\it after the fact} what dynamical quantities (e.g., orbital elements) to extract.
An important difference in this data analysis stage is that (unlike with the original simulation) the task can be easily parallelized.
We have implemented these features in the open-source N-body package \reb.

Although, for the sake of definiteness, we focus on the Solar System and the HL~Tau system in this paper, our discussion is equally applicable to a wide variety of dynamical systems.
The paper is organized as follows.
We first discuss in detail the issue of reproducibility in~Sec.~\ref{sec:reproducibility}.
Then, we describe the concept of the \sima and how it can be used to efficiently analyze long-term simulations in~Sec.~\ref{sec:simulationarchive}.
We look at two examples in~Sec.~\ref{sec:examples}, a long term integration of the Solar System (Sec.~\ref{sec:solarsystem}) and an unstable system resembling the orbital parameters of the HL~Tau system (Sec.~\ref{sec:hltau}).
We conclude with a summary in~Sec.~\ref{sec:conclusions}.

\section{Barriers to reproducibility}
\label{sec:reproducibility}
Current N-body simulations, especially those of chaotic systems, are only reproducible in a statistical sense.
Different groups employing the same simulation will only agree on the {\it distributions} of outcomes from a large enough number of simulations.
However, one is often interested in particular outcomes, e.g., Mercury diffusing onto a collisional trajectory with Venus \citep{LaskarGastineau2009}, or Jupiter ejecting a fifth giant planet from the Solar System to explain the current orbital architecture of the giant planets \citep{Nesvorny2012}.
In such a case, two groups will not be able to reproduce the same outcome even if they share their exact initial conditions.
To obtain these rare cases, one would have to expend the computational cost of a large suite of simulations to extract a few qualitatively similar but distinct trajectories.
This is not only grossly inefficient, but also renders detailed comparisons impossible.

One of the first obstacles in reproducibility of numerical simulations is the lack of access to source code \citep{Baker2016}.
Although many authors have published their numerical methods not only as papers but with the accompanying code \citep{Chambers1997, Duncan1998, ReinTamayo2015}, others have not. 
Whereas one should in principle be able to reconstruct the algorithm from a description in a paper, this is often not practical, and is effectively impossible when the goal is bit-wise reproducibility.
As a simple example, consider the expressions
\begin{eqnarray}
y_1 = (2/3)\cdot x\quad\quad \text{and}\quad\quad y_2 = (2\cdot x)/3\nonumber
\end{eqnarray}
which are mathematically equivalent but will yield different results on a computer with finite floating point precision\footnote{Note that $y_2$ is in general the better choice because it randomizes errors. By contrast, $y_1$ might lead to biased results because $2/3$ evaluates to the same rounded number each time the expression is evaluated.}.
Thus, without access to the actual source code used in a simulation, there is no hope of achieving bit-wise reproducibility.

Unfortunately, it is often overlooked that even the exact source code (or even binary executable) and initial conditions might produce distinct outcomes on different machines.
Machine-dependent results can originate from physical differences in hardware, compilers, compiler options, operating systems or libraries.
In fact, only a very limited set of floating point operations such as additions and multiplications are guaranteed by the IEEE 754 floating point standard \citep[see][]{IEEE754} to give the same results on all machines that comply with the standard. 
We note in particular that often-used functions such as \texttt{sin()}, \texttt{cos()}, and \texttt{pow()}, and thus any codes that employs them, are {\it not} machine independent. 

Finally, one obviously requires a set of initial conditions to reproduce an N-body simulation.
Bit-wise reproducibility additionally demands storing these initial conditions in a binary format, in order to have the exact floating point representation of all numbers; a text-based format is not sufficient.

\section{The Simulation Archive}
\label{sec:simulationarchive}
We now describe how to overcome the barriers to reproducibility mentioned above. 
First, it is essential to work with open-source software and tools.
We work with \reb \citep{ReinLiu2012}, and in particular the integrators \whfast \citep{ReinTamayo2015} and \ias \citep{ReinSpiegel2015}. 
\whfast is a heavily optimized implementation of the standard second-order Wisdom-Holman algorithm \citep{Kinoshita1991, WisdomHolman1991} and can be used with symplectic correctors up to order 11 \citep{Wisdom2006} in both Jacobi and heliocentric coordinates.
\ias is a high-order, adaptive Gauss-Radau scheme that is accurate to machine precision and can handle both conservative and non-conservative forces.
Both the symplectic \whfast and non-symplectic \ias integrators are machine independent. 

We achieve this by substituting function calls to mathematical libraries with appropriate series expansions.
For example, we implement a series expansion of Stumpff functions for \whfast \citep{ReinTamayo2015}.
For \ias, we implement our own routine to calculate $7$th roots in order to avoid calling the machine-dependent  \texttt{pow()} function.
Implementing our own mathematical functions often also leads to a more accurate and faster algorithm \citep{ReinTamayo2015}.
We also note that \reb is written in C99, a programming standard that explicitly states that floating-point operations may not be rearranged by the compiler unless it can guarantee that the result is bit-wise equivalent. 
This is still true if optimization flags such as \texttt{-O3} are used, but not if \texttt{--fast-math} is enabled. 

Thus, by using \reb and one of the above integrators, simulations are automatically machine independent. 
If authors choose to share their initial conditions, simulations are then in principle bit-wise reproducible. 
However, several additional technical details have to be overcome in order to replicate the same series of timesteps.
In the following section we describe our concept of the \sima, which addresses these issues, and allows users to recreate and restart integrations at arbitrary times almost instantaneously.
This enables exciting new ways to analyze and share simulation data.

\subsection{A single binary file}
General purpose binary file formats for $N$-body simulations have already been developed by several other groups \citep[see e.g.][]{Faber2010,Farr2012,Cai2015}.
This paper is not about a new binary format for $N$-body simulations. 
Rather, we describe how to store data in a way that allows simulations to be restarted in a way that reproduces the original simulation exactly.
To our knowledge, this has never been done before in a publicly available code that is used for long term planetary simulations.

What we refer to as the \sima is a single binary file that includes all relevant information to reproduce an $N$-body simulation down to the last bit.
In particular, we store:
\begin{itemize}[leftmargin=*]
    \item[-] Name and version of the program used to generate the file
    \item[-] Initial conditions
    \item[-] All constants and configuration parameters, such as gravitational constant and timestep
\end{itemize}
Then, as the simulation progresses, we append the binary file with snapshots of all data needed to restart the simulation at that time.
In the case of a simulation of the Solar System with eight planets, we typically append a snapshot every $50\,000$~yrs to the \sima.
Each snapshot is only 500 bytes long but contains all the necessary information to exactly restart a simulation at that time.
For a $10$~Gyr integration, the total file size of the \sima is about~100~MB.
Cosmological or other large-N simulations would have to find a different balance between file size and the computation time between snapshots.

We note that our implementation does currently not support big-endian machines.
However, since all modern x86/x64 architectures are little endian, we see few reasons to implement this feature.
Should the need arise, it is straightforward to extend the format to be endian agnostic.

Furthermore, we note that our implementation supports both 32 and 64~bit architectures. 
However, files created on 32~bit machines can currently not be read on a 64~bit machine and vice versa.
Given the widespread adoption of 64~bit systems these days, we see few reasons to implement the \sima in an architecture agnostic way.
Should the need arise, this feature is straightforward to implement and does not affect the reproducibility of simulations.

\subsection{Extracting arbitrary quantities at any time}
Many N-body codes allow one to save binary snapshots of simulations.
The \sima goes beyond this by guaranteeing that different machines reloading and integrating the same snapshot will give identical results bit-by-bit.

The storage requirements for one snapshot are very small, only 500~bytes for a \whfast simulation of the Solar System.
This allows us to store many snapshots which are typically spaced only a few seconds apart in wall time.
One can therefore access any value of any particle at any timestep by loading the snapshot just prior to the requested time and then integrating the simulation forward in time. 
Since restoring a simulation from a binary file is almost instantaneous and snapshots are only a few seconds apart, one has to wait at most a few seconds for this short integration to finish and yield the desired quantities.
Furthermore, we note that this process is easily parallelizable if one wants to access parameters at multiple times (See Sec.~\ref{sec:solarsystem} and \ref{sec:collision}).

We now address technical details pertaining to the class of mixed-variable symplectic integrators \citep[e.g.,][]{SahaTremaine1992}, which are the de facto standard for long-term planetary integrations, to non-symplectic integrators, and to the incorporation of additional effects beyond point-source gravitational forces.

\subsection{Mixed-variable symplectic integrators and synchronization}
When trying to restart a simulation using a mixed-variable symplectic (MVS) integrator like \whfast, one encounters three synchronization issues.
We detail how we overcome these challenges in \whfast specifically, but emphasize that these pitfalls and solutions are applicable to MVS integrators in general.

First, MVS integrators split the evolution of the system into exactly solvable steps, which are then interleaved with one another.
The \whfast integrator, like many other MVS integrators, is a leapfrog-style drift-kick-drift scheme.
The drift step corresponds to an operator evolving the system under the Keplerian part of the Hamiltonian and the kick step evolves the system under the interaction term. 
Each of the two drift steps evolves the system for half a timestep. 
In a long simulation, one can speed up the calculation by combining the last and first drift steps of adjacent timesteps into one.
However, if positions and velocities are required after a particular timestep, one must reintroduce the half-timestep drift step, a process that we refer to as synchronizing the positions and velocities. 
This leads to a problem if one considers two integrations of the same trajectory across a particular timestep, where in one case one chooses to output particle information (and thus must synchronize), and in the other one does not (and thus performs a full drift step).
While evolving the system under the drift operator for two half-timesteps (in order to output data) is mathematically equivalent to a full drift step, the two will give a different numerical result because of the finite floating point precision.
To deal with this, the \sima always stores the particle positions and velocities in the unsynchronized state following the kick step. 
If an additional half drift step is required to synchronize the positions and velocities, the result is then discarded before the integration continues with a full drift step from the state that was cached by the \sima.

Second, the drift and kick steps are often solved in different coordinate systems \citep[e.g.,][]{WisdomHolman1991}.
The Keplerian part is best solved in Jacobi coordinates\footnote{Jacobi coordinates are advantageous for systems with planets on well separated orbits such as the Solar System. For other systems where orbits might cross, heliocentric coordinates might be beneficial.},
whereas the interaction Hamiltonian is solved in an inertial frame.
We thus have to convert back and forth between two different coordinate systems, which can harm precision in long term integrations \citep{ReinTamayo2015}.
However, at closer inspection one finds that the interaction step only needs to access the positions in the inertial frame, not the velocities.
Furthermore, only the accelerations calculated in the inertial frame are needed in Jacobi coordinates.
Thus, a simulation can stay in Jacobi coordinates for the entire integration and only calculate the positions in the inertial frame for the interaction step (but not recalculate the Jacobi coordinates from those positions as positions do not change in the interaction step).
In the \sima, we store the Jacobi coordinates, not the coordinates in the inertial frame. 
Were we to store the inertial coordinates and then recalculate the Jacobi coordinates later, we would, once again, get a different numerical result because of finite floating point precision.
Thus, we would not be able to restart simulations exactly.

Third, when symplectic correctors are used, a corrector step is performed at the beginning of the integration and whenever an output is needed \citep[e.g.,][]{Wisdom2006}.
Similar to the calculation from Jacobi coordinates to the inertial frame and back, this corrector step can degrade the precision and speed in long term simulations if applied repeatedly.
And again similar to above, we would not be able to restart a simulation exactly were we to store the coordinates after applying a symplectic corrector.
This is why when storing coordinates in the \sima, we do not apply the symplectic corrector steps beforehand.
The corrector step will be performed at a later time.

In summary, in the \sima, we store the Jacobi coordinates of all particles in what we call an unsynchronized state, i.e. after the kick step and without applying symplectic correctors. 
Only the combination of these concepts allows us to restart simulations exactly down to the last bit.

As a result, if we want to get any physical meaningful output from the \sima, we need to synchronize the coordinates beforehand. 
To reemphasize, this is done at a time the simulation is analyzed, not while the simulation is running.
In this synchronization step, we first perform half a drift step to advance positions and velocities to the same time. 
We then apply the symplectic correctors and convert from Jacobi to inertial coordinates.

If we want to restart a simulation (and do not require an output), then we do not need to perform the synchronization steps and simply restart from the Jacobi coordinates in the unsynchronized state.

We also implement the \sima for the heliocentric version of \whfast.
Heliocentric coordinates can be beneficial for system in which orbits are crossing \citep[for a discussion of differnet coordinate systems, see e.g.][]{HernandezDehnen2016}.
The synchronization issues described above are exactly the same, except that we have heliocentric coordinates instead of Jacobi coordinates.
Indeed, the above considerations need to be addressed to create any reproducible MVS integrator.

\subsection{IAS15}
For integrators that do not rely on coordinate transformations restarting simulations bit-by-bit is more straightforward.
However, some issues might still arise, for ecample if the integrators depends on information from previous timestepssuch as a predicted values for the timestep and coordinates. 
We implemented the \sima for the \ias integrator \citep{ReinSpiegel2015}.
For this specific integrator, one needs to store several internal arrays which \ias uses to predict values for the next step and speed up convergence. 
Because an iteration in \ias might converge to a slightly different value depending on the initial guess, we need to store these predicted values at every snapshot to guarantee bit-wise reproducibility.
This results in a single snapshot being about a factor of 10 larger than for \whfast.

\subsection{Additional forces and post-timestep modifications}
As discussed above, when accessing a \whfast \sima snapshot to output physically meaningful quantities, we have to perform a synchronization step. 
This can lead to a further issue if non-gravitational effects are present, which is often the case in $N$-body simulations, for example to model the effects of general relativity or tides. 

In \reb these effects are modelled as either additional forces or so called post-timestep modifications.
If the effects are simply forces, then they do not need to be applied for the synchronization step which only involves a drift step, but not a kick step (where forces are calculated). 
However, if post-timestep modifications are used, then these need to be also applied when the synchronization steps are performed.
Furthermore, if symplectic correctors are used, then either kind of effect must be present at the time of synchronization because symplectic correctors apply both the kick and drift steps repeatedly.

\section{Examples}
\label{sec:examples}

We now give two concrete examples of how the \sima can be used in practice, one with the symplectic \whfast algorithm, and the other with the adaptive, high-accuracy \ias integrator.

\subsection{Solar System}
\label{sec:solarsystem}
As an example, we now integrate the eight major planets of our Solar System forward in time. 
The same discussion applies to other dynamical systems such as exoplanetary systems.

The shortest timescale in the Solar System that we consider is the 88-day orbital period of Mercury.
The longest timescale we are interested in is roughly the age of the Solar System, 4.6~billion years, which also happens to approximate the remaining lifetime of the Sun on the main sequence.  
The huge separation of scales (11 orders of magnitude) makes this an extremely difficult problem to calculate, even though the physical laws governing the motion of planets have been well known since Newton (ignoring small general relativistic and other corrections). 

Because of this, the dynamical evolution of the Solar System still offers many surprises for researchers to this day.
In fact, only in the last 25 years have direct numerical integrations of all Solar System planets become feasible.
This is mainly thanks to the development of fast computers and novel integration methods, most notably mixed-variable symplectic integrators \citep{Kinoshita1991, WisdomHolman1991}.
Nevertheless, accurate direct numerical integrations of the eight Solar System planets over several billion years may still require months of computing time per simulation and several million CPU-hours for an ensemble of simulations.
Attempts to significantly speed up simulations with the help of parallelization are futile as the underlying problem is inherently sequential \citep[although attempts have been made, see][]{Saha1997}.

A further complication arises because the Solar System is chaotic on timescales of several million years, whereas the life-time of the system is several billion years.
One thus has to integrate an ensemble of simulations to draw statistical conclusions of the system's long-term evolution.
For example \cite{LaskarGastineau2009} required more than $6$~million hours of wall time for a set of 2500 simulations, each spanning 5~Gyrs.

\subsubsection{Initial conditions and integrator}
For our integration of the Solar System, we use the NASA Horizons system to provide us with initial conditions\footnote{\url{http://ssd.jpl.nasa.gov/?horizons}}.
More accurate ephemerides are available \citep{Fienga2011}, but given the other approximations we make (see below), we feel confident that the initial conditions we use are accurate enough to capture the fundamental physical effects.

We use the symplectic Wisdom-Holman type integrator \whfast \citep{ReinTamayo2015} mentioned above with a timestep of $6$~days. 
To improve the energy conservation, we use a symplectic corrector of order~11 \citep{Wisdom2006}.
With these settings, a simulation takes approximately one month of wall time on an Intel Xeon CPU (E5-2697 v2, 2.70GHz).

\subsubsection{Non-gravitational effects}
We include a post-Newtonian correction for general relativity in the form of a simple potential \citep{Nobili1986},
\begin{eqnarray}
    \Phi_{\rm GR} &=&  -\frac{3 G^2 m M_\odot^2}{c^2} \frac1{r^2}
\end{eqnarray}
for each planet with mass $m$. In the above equation $M_\odot$ is the mass of the Sun, $r$ is the heliocentric distance of the planet, and $G$ and $c$ are the gravitational constant and the speed of light, respectively.
This ensures that we reproduce the correct apsidal precession frequencies for the planets, in particular that of Mercury.
We ignore all other non-gravitational effects and the Earth-Moon system is treated as one particle.

\subsubsection{Simulation archive}
We use the \sima for our integration.
Individual snapshots are $51\,434$~years apart, which corresponds to roughly $10$~seconds in wall time.
We integrate the system for $10$~billion years, thus ending up with $194\,000$ snapshots. 
Given that each snapshot requires $500$~bytes, this yields a final \sima binary file of roughly $100$~MB.

With this setup, we can now access an arbitrary time during the entire $10$~Gyr integration within 5~seconds on average and within at most 10~seconds.

\subsubsection{Results}

\begin{figure}
 \centering \resizebox{\columnwidth}{!}{\includegraphics{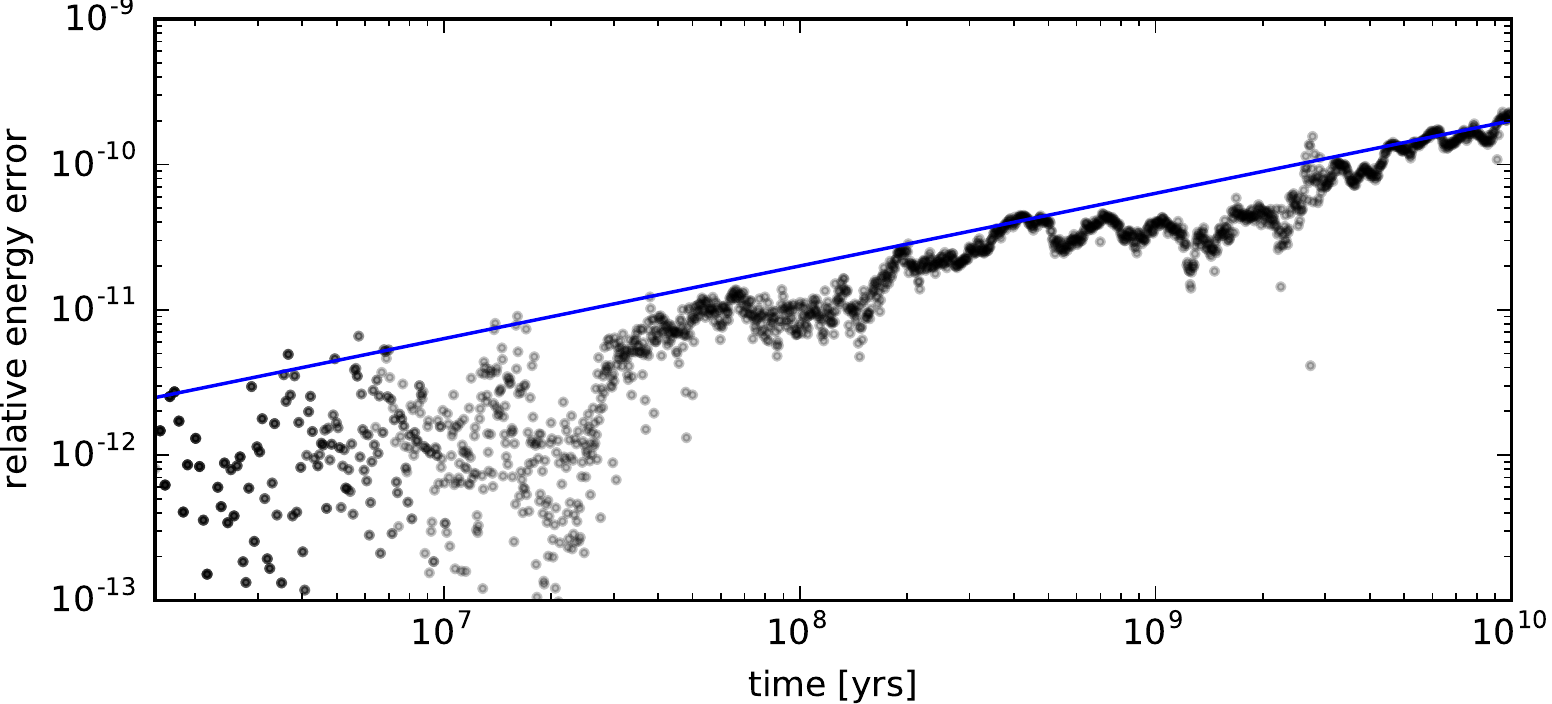}}
 \caption{
     This plot shows the relative energy error of our Solar System simulation as a function of time. 
     The blue curve shows the slope expected from an unbiased ~$t^{1/2}$~error growth.
     It takes 1.9~seconds to generate the data for this plot from the \sima and then render the plot.
\label{f_energy}}
\end{figure}

We have run several hundred simulations of the Solar System over $10$~Gyrs with slightly different initial conditions (fractional differences of $10^{-10}$).
We will present the full set of simulations and a discussion of the physical results in a future paper.
In this paper, we only use one simulation in the ensemble to demonstrate the \sima.

In Fig.~\ref{f_energy} we show the relative energy error of the system as a function of time. 
The straight blue line corresponds to a sub-linear error growth $\propto~t^{1/2}$. 
We can see that the error growth follows this $t^{1/2}$ behaviour for the entire $10$~Gyr integration, thus confirming that the \whfast integrator is unbiased and follows Brouwer's law \citep{Brouwer1937,ReinTamayo2015}.
We created this plot using the snapshots stored in the \sima.
This kind of analysis is extremely fast. 
Reading the binary file, restoring the simulations, synchronizing the integrator, and calculating the energy error for every snapshot, followed by producing the plot, takes less than 2~seconds.

\begin{figure}
 \centering \resizebox{\columnwidth}{!}{\includegraphics{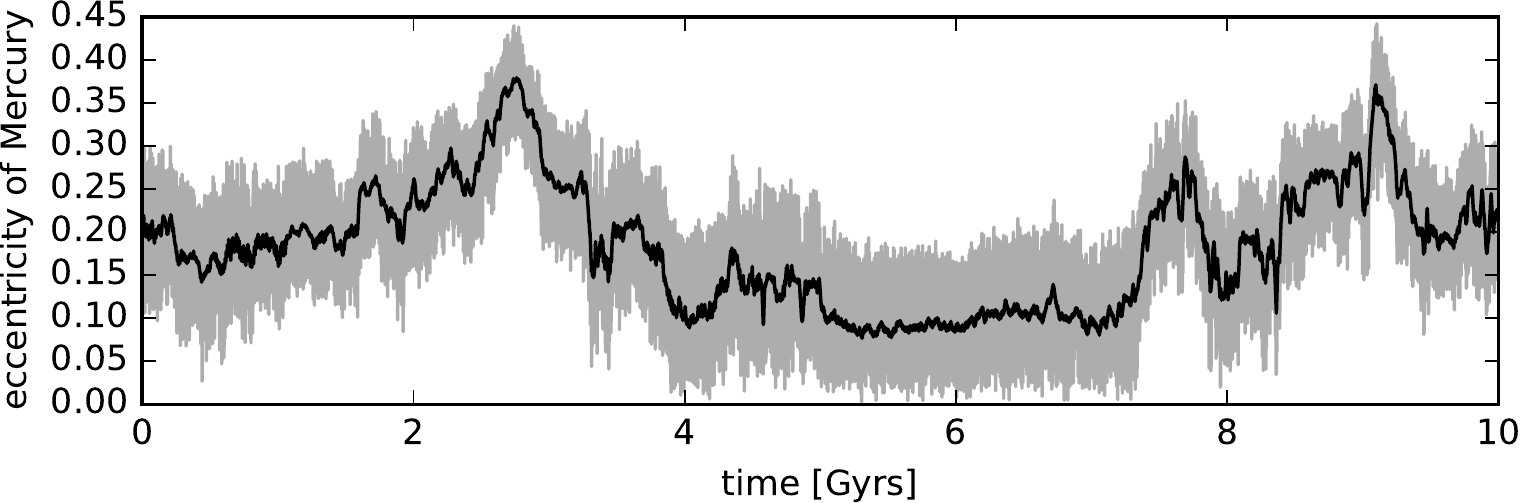}}
 \caption{
     The gray line shows the eccentricity of Mercury as a function of time.
     Because the eccentricity behaviour is dominated by short-term oscillations, we also plot a moving average with a 1.5-million-year window as a black line. 
     It takes 2.3~seconds to generate the data for this plot from the \sima and then render the plot.
\label{f_eccentricity}}
\end{figure}
In Fig.~\ref{f_eccentricity}, we plot the eccentricity of Mercury as a function of time (gray curve). 
We also over-plot the running average over a window of $1.5$~Myrs as a black curve to filter out high-frequency oscillations.
This plot illustrates that Mercury remains stable in this particular simulation as expected for $\approx99$\% of the simulations \citep{LaskarGastineau2009}.
Again, we used the \sima to produce this plot after the simulation had finished.
Reading the binary file, restoring the simulations, synchronizing the integrator, calculating the orbital elements, calculating the running average and producing the plot takes only about 2~seconds.

\begin{figure}
 \centering \resizebox{\columnwidth}{!}{\includegraphics{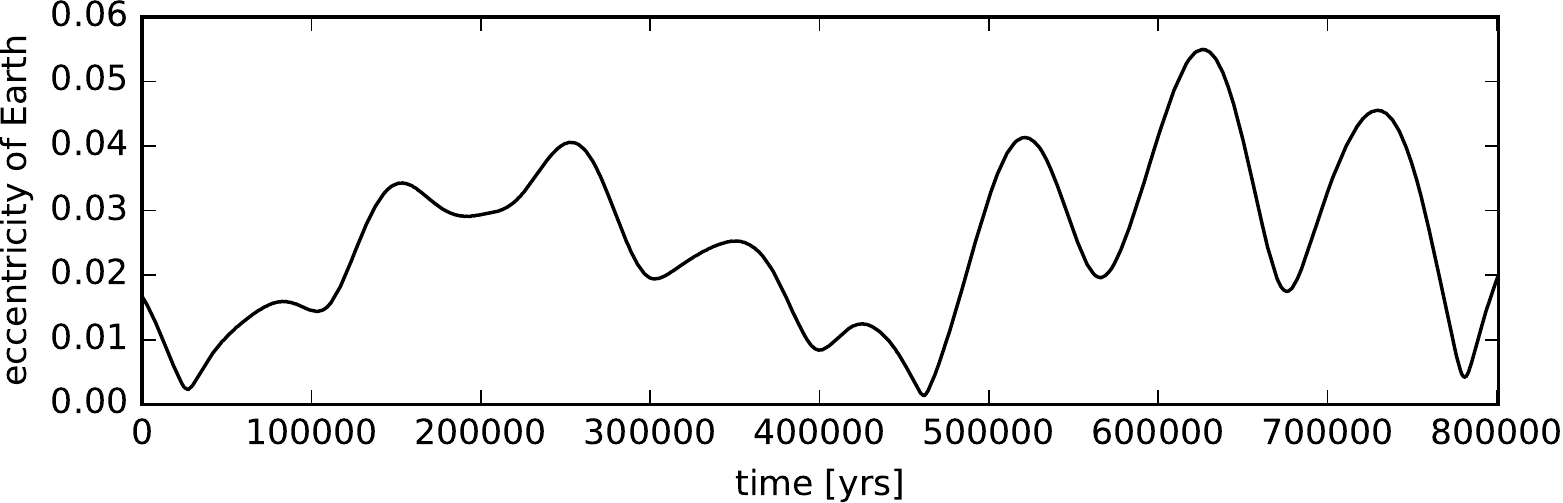}}
 \centering \resizebox{\columnwidth}{!}{\includegraphics{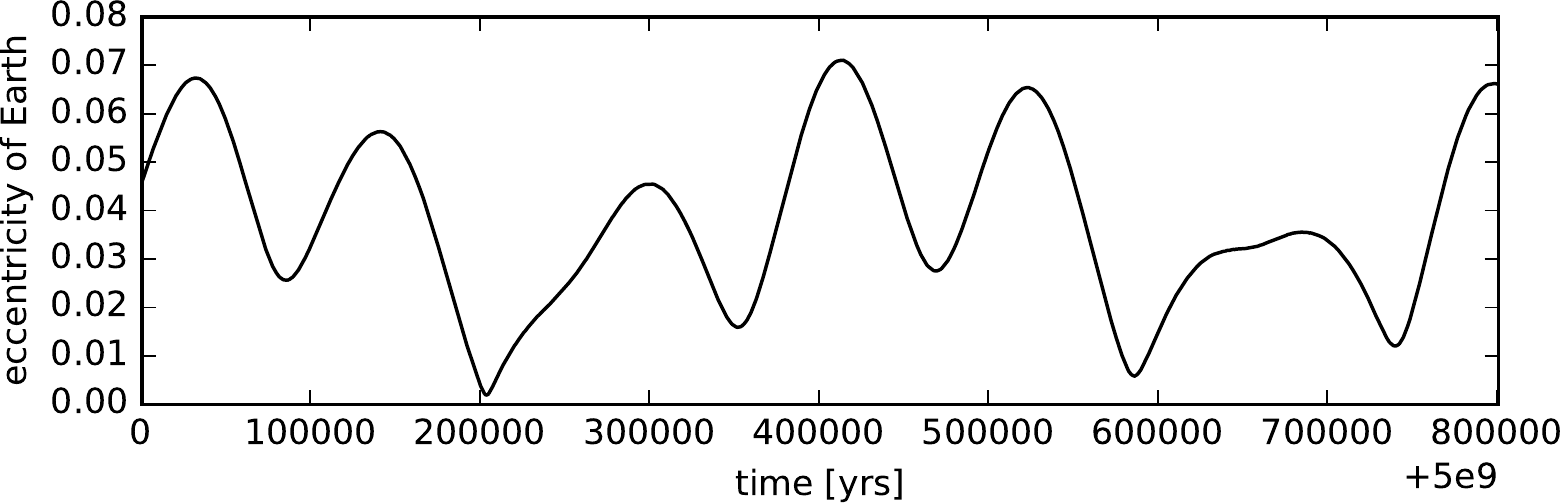}}
 \caption{
     The top and bottom panel both show the eccentricity of the Earth over a $800\,000$~year period. 
     The top panel starts at the present day, while the bottom panel starts 5~billion years from now. 
     The oscillations visible in these plots are related to Milankovitch cycles.
     It takes only 41~seconds to generate the data for this plot from the \sima in parallel with 24 threads and then render the plot.
\label{f_eccentricity_earth}}
\end{figure}
The first two plots illustrate the usefulness of the \sima in quickly analyzing and visualizing very long integrations.
The next figure illustrates how to extract data at a higher cadence than provided by the snapshots in the \sima.
Figure \ref{f_eccentricity_earth} plots the eccentricity variations of the Earth over a $800\,000$~yrs interval, starting at the present day (top panel) and 5~billion years from now (bottom panel).
One can see large oscillations on a roughly $100\,000$~yr timescale, which are associated with Milankovitch cycles and changes in the Earth's climate. 
Because the variations happen on a timescale comparable to the output cadence in the \sima, we cannot rely on the snapshots in the \sima to generate this plot.
Instead, we load a snapshot that is nearby and then re-integrate the simulation forward in time, this time outputting data more frequently.
The entire process is automated so that a user can simply request a simulation at a given time, and the \sima will load the nearby snapshot and integrate it to that time.
Note that this process is trivially parallelizable, allowing us to speed up the analysis by making use of multi-core desktop machines and clusters. 
While producing Fig.~\ref{f_eccentricity_earth}, we made use of 24 threads, rendering both plots in only 41~seconds.
The bit-by-bit reproducibility of the \sima means that we are no longer restricted to a predefined output cadence; we can always resample the simulation after the fact to investigate interesting features.

\subsection{Collisional trajectory}\label{sec:collision}
\label{sec:hltau}
\begin{figure*}
 \centering \resizebox{\textwidth}{!}{\includegraphics{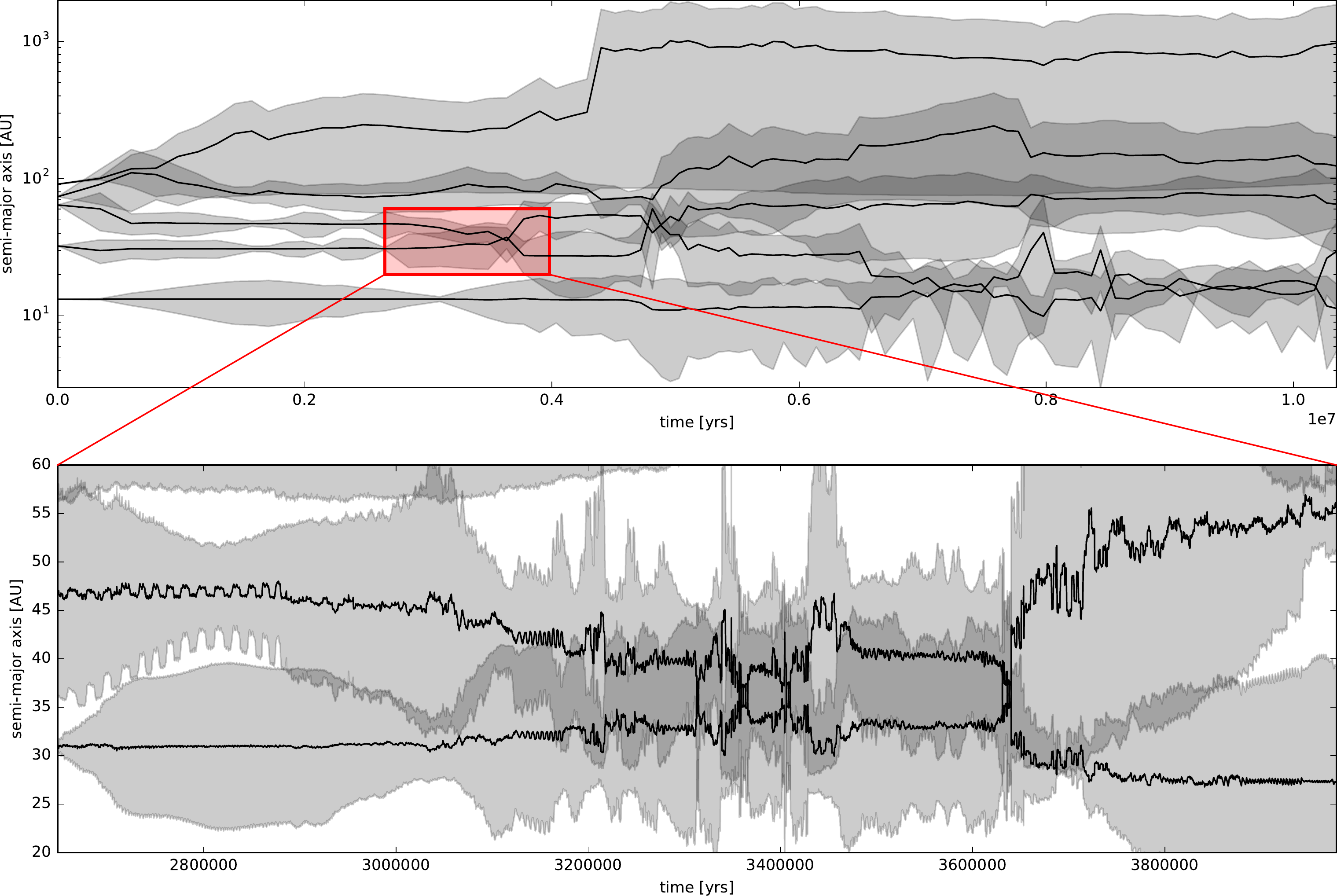}}
 \caption{
     The semi-major axes (solid black lines) of the five planets in a hypothetical HL Tau system. 
     The gray areas correspond to the radial range between the periastra and apastra of the planets. 
     Planets two and three have a close encounter after $\sim 3.5$~Myrs.
     To analyse this specific close encounter, we restart the simulation just before the close encounter and increase the output cadence to generate the bottom plot. 
     Thanks to the exact reproducibility of the \sima, the same close encounter is reproduced.
\label{fig:encounter}}
\end{figure*}

The \sima can also be used to analyze systems with close encounters, for example if studying the ejection of a fifth giant planet from the Solar System \citep{Nesvorny2012}.
As another example, we here consider an integration of five planets in the HL Tau system, with semi-major axes at the locations of the five most prominent gaps in the circumstellar disk as reported by \cite{Brogan2015}.
We assign the star one solar mass, and the planets a mass $3\cdot10^{-4}$ times smaller, roughly one Saturn mass, which will quickly lead to crossing orbits and dynamical instability \citep{Tamayo2015}.
The orbits are initialized as circular, with an inclination drawn from a uniform distribution between 0 and 1 degrees.
The remaining angles are drawn randomly from a uniform distribution.

We integrate the system for 10 million years using \ias, which is ideally suited to study such an unstable system, given its adaptive timestep.
For such unstable systems, where equal intervals in simulation time might correspond to vastly different intervals in computation time while resolving particularly close encounters, it is useful to specify the output cadence in wall time.
This ensures that integrating from one snapshot to the next always takes an equal amount of time.
In our case, we save snapshots at 10~second intervals in wall time, yielding a binary of $\sim 3$~MB.

In Fig.~\ref{fig:encounter}, we plot the semi-major axes of all planets as solid black lines, and also show the radial range between their periastra and apastra shaded in grey.
The initial output sampling in the top plot is coarse, but we can easily identify that the second and third planets had a close encounter after $\sim3.5$~Myrs.
To further analyze this encounter, we restart the simulation just before the encounter and increase the output cadence.
This is only possible because the bit-by-bit reproducibility ensures that the exact same close encounter will occur in the restarted simulation.
To our knowledge this would not currently be possible with any other publicly available integrator.

\section{Conclusions}
\label{sec:conclusions}
We present a new paradigm for performing $N$-body simulations that allows researchers to reproduce one another's results exactly.
This is fundamentally different to what has been done before, where, in the best case, simulations were only reproducible in a statistical sense.
We believe this will improve the scientific method's ability to operate effectively.

We achieve bit-by-bit reproducibility by carefully implementing machine-independent integrators and storing all initial conditions and parameters in a binary file.
Because our approach allows everyone (not just the original authors of a study) to reproduce particular simulations exactly, this opens up new possibilities for sharing and analyzing dynamical systems.
In particular, this makes it possible for the first time to reanalyse particular trajectories in chaotic systems that lead to interesting (and perhaps rare) outcomes which otherwise would be unrecoverable.

We describe the concept of our \sima, which enables the above and provides an easy-to-use interface for extracting simulation data at saved snapshots, or at intermediate times.
This makes it easy and efficient to analyze the simulation retrospectively, and to output arbitrary quantities, exactly (bit-by-bit) as if one had stored them in the original run. 
Because our algorithms are machine independent, \simas can be shared between groups, even if they use different operating systems, compilers or libraries.

We hope that the ideas presented in this paper will become a new paradigm for running $N$-body simulations.
This includes the expectation that when scientists present N-body simulations, that they would share their machine-independent source code, as well as their exact initial conditions and all other numerical parameters and constants.
Such a practice would make it possible to verify and extend one another's results, rendering the scientific enterprise more efficient and robust.

The \sima has been implemented in the \reb code which can be downloaded at \url{https://github.com/hannorein/rebound}. 
iPython notebooks to reproduce the plots in this paper can be downloaded at \url{https://github.com/hannorein/reproducibility-paper}.

\section*{Acknowledgments}
This research has been supported by the NSERC Discovery Grant RGPIN-2014-04553.
We thank an anonymous reviewer for valueable feedback that helped us improve the manuscript.

\bibliography{full}
\end{document}